# One-Two Dimensional Nonlinear Pulse Interaction

A. Ciattoni[1,4], A. Degasperis[2] and E. DelRe[3,4]

[1]Dipartimento di Fisica, Universita' dell'Aquila,67010 L'Aquila, Italy
[2]Dipartimento di Fisica, Universita' di Roma "la Sapienza,"00185 Roma, Italy and Istituto Nazionale di Fisica Nucleare, Sezione di Roma1, Roma, Italy
[3]Fondazione Ugo Bordoni, Via Baldassarre Castiglione 59, 00142 Roma, Italy
[4]INFM, Unità di Roma I, Italy

**Abstract**

The peculiar intergrability of the Davey-Stewartson equation allows us to find analytically solutions describing the simultaneous formation and interaction of one-dimensional and two-dimensional localized coherent structures. The predicted phenomenology allows us to address the issue of interaction of solitons of different dimensionality that may serve as a starting point for the understanding of hybrido-dimensional collisions recently observed in nonlinear optical media.

Nonlinear wave propagation occurs in many different physical systems (e.g. water waves and optics) and leads to a myriad of different interesting and useful phenomena that are in striking contrast to linear propagation effects [1]. Among these, the emergence of localized non-dispersive coherent pulses, solitons, that do not suffer deformation and that undergo elastic-like collisions, have attracted much attention [2]. Perhaps their most peculiar physical feature is associated with their interaction dynamics. Whereas classical soliton experimental studies have been confined to lower dimensional (one-dimensional, 1+1D) systems, where diffraction occurs in one dimension only, in the last decade experiments in nonlinear optics have allowed the stable observation of both 1+1D solitons in a bulk environment, known as stripe or wall solitons, and two-dimensional (2+1D) needle solitons, where linear deformation is halted in two pulse dimensions [3]. Surprisingly, such phenomena can be observed in the same nonlinear medium *simultaneously*, and have permitted the observation of a new solitonic process: the collision and interaction of two solitons of different dimensionality, one being a needle, the other a stripe, in a photorefractive crystal [4]. Furthermore, recent experiments in near-resonant gases have allowed the study of interaction between a vortex and a dark stripe soliton [5]. To our knowledge a stripe-needle collision has never been theoretically investigated, nor in optical physics, or in any other context for that matter. The description, apart from encountering the "standard" difficulties connected to nonlinear partial differential equations (only rarely integrable), poses a number of modeling riddles. To name one, the model should support both stripe and needle solutions, and most importantly, needle-stripe hybrid solutions, a circumstance that even in its linear realization poses peculiar issues [6]. Even more, although a numerical investigation of a Kerr-saturated model has been performed [4], in order to obtain a clear and complete picture, we would like an actual integrable nonlinear model. In this Letter we tackle the hybrid collision in the frame of the Davey-Stewartson equation (DS) [7], a generalization of the nonlinear Schroedinger equation (NLS) [8], that is known to allow for the explicit analytical description of stripe solitons *and* needle solitons, *separately* [9]. After showing that the system can actually support a hybrido-dimensional structure, we are able to derive *fully-analytical* solutions for the needle-stripe interaction, making use of the "dressing" theorem [10]. For zero-angle collisions we find periodic "breathing", hinting at the possible existence of stationary hybrid states. For angled interaction, we find that the two components, needle and stripe, retain their original identity and localization after the collision, even though the needle changes shape.

The DS equation is a generalization of the NLS equation

$$iQ_z + Q_{xx} + c|Q|^2 Q = 0 \ , \tag{1}$$

where the binding self-interaction potential $-c|Q|^2$ (c being a positive parameter) is *local* and responsible for the existence of localized, stable, nondispersive pulses $Q(x,z)$. DS solitons are solutions of the DS equation

$$iE_z + E_{xx} + E_{yy} + V E = 0 \ , \tag{2a}$$

$$V(x,y,z) = v(x,z) + u(y,z) + (1/2)\left[\int_{-\infty}^{x} (|E(s,y,z)|^2)_y ds + \int_{-\infty}^{y} (|E(x,s,z)|^2)_x ds\right] \tag{2b}$$

where the binding potential $V(x,y,z)$ is now a *nonlocal* expression of the intensity $|E|^2$, and v and u are two given arbitrary binding one-dimensional (i.e. v on the x-axis and u on the y-axis) potentials ("waveguides") which are responsible for the formation of localized pulses $E(x,y,z)$. These would exist even if the self-interaction terms were *not* present (say, in the linearized limit of eq.(2)). Thus, in the two-dimensional case, solitons are formed by "external waveguides", whereas the interaction is mediated by the nonlinear terms. To emphasize this difference with respect to NLS solitons, DS solitons are sometimes referred to as "dromions" [11]. As opposed to the full two-dimensional localization, in the important case of vanishing external potentials, v = 0 and u=0, solitons can be localized in one single direction, say in the transverse coordinate $T = x\cos\vartheta + y\sin\vartheta$. These solutions of eq.(2), $E = E(T,z)$, are actually also solutions of eq.(1) with $c = 1/\sin(2\vartheta)$ for $0 < \vartheta < \pi/2$, and represent straight walls in the (x,y) plane. This equivalence can be used to identify the physical lengths involved in obtaining dimensionless variables throughout the paper, namely the dispersion length and the nonlinear length [8]. Note that NLS soliton collisions [12] and dromion collisions have been extensively investigated by means of multi-soliton solutions, their differences being well-known [9].

In terms of the DS equation (eq.(2)), the hybrid collision under study is that of a wall soliton and a needle dromion. Taking advantage of the fact that the (one-dimensional) NLS solitons appear themselves as wall solutions in the (x,y) plane of the DS equation (as pointed out above), we consider exact solutions of eq.(2) which are the nonlinear superposition of a wall (NLS 1-D soliton in the transverse coordinate T) and a ball (dromion).

A variety of solutions of the DS equation have been already constructed, including the one-wall solution, using various techniques, such as Backlund transformations [13], bilinearization [14] and inverse spectral method [9]. Our approach here is based on a dressing formula [9] [10], which follows from the spectral transform method of constructing solutions which vanish at infinity in all directions (say as $x^2 + y^2 \to \infty$) but whose validity can be easily recognized to include also wall-type solutions. This formula (for details and generalizations, see [11]) reads

$$E = A_{11} f^{(1)} g^{(1)} + A_{12} f^{(1)} g^{(2)} + A_{21} f^{(2)} g^{(1)} + A_{22} f^{(2)} g^{(2)} \qquad (3)$$

where $f^{(j)}(x,z)$ and $g^{(j)}(y,z)$, j=1,2, are solutions of the (linear) Schroedinger equations $if^{(j)}_z + f^{(j)}_{xx} + uf^{(j)} = 0$ and, respectively, $ig^{(j)}_z + g^{(j)}_{yy} + vg^{(j)} = 0$, and the functions $A_{jn}(x,y,z)$ are the entries of the 2x2 matrix

$$\mathbf{A} = 2\mathbf{R}( \mathbf{1} + \mathbf{G} \mathbf{R}^+ \mathbf{F}^* \mathbf{R} )^{-1} , \qquad (4)$$

$\mathbf{R}$ being a diagonal constant matrix, $R_{jn} = a_j \delta_{jn}$, where its entries $a_1$ and $a_2$ are two complex parameters, while the Hermitian matrices $\mathbf{F}(x,z)$ and $\mathbf{G}(y,z)$ are defined by the integral expressions

$$F_{jn} = \int_{-\infty}^{x} f^{(j)}(s,z) f^{(n)*}(s,z)\, ds , \quad G_{jn} = \int_{-\infty}^{y} g^{(j)}(s,z) g^{(n)*}(s,z)\, ds . \qquad (5)$$

The linear approximation of eq.(3), which is obtained by replacing the matrix **A** with 2**R**, is just the linear superposition of the two solutions $f^{(j)}(x,z)g^{(j)}(y,z)$, j=1,2, of the linearized DS equation (that is the Schroedinger equation with a separable potential). Furthermore, the expression which is obtained by setting in eq.(3) $a_2 = 0$ ($a_1 = 0$) is a solution of eq.(2) and, throughout the paper, we associate with j=1 the wall solution and with j=2 the dromion (ball) solution. With this terminology, therefore, eq.(3), with definitions of eqs.(4) and (5), describes the interaction of a wall and a ball, and our task is to display the main properties of this collision by analyzing both graphically and analytically this formula.

To highlight the relevant aspects of the hybrid nonlinear interaction we assume that v=v(x) and u=u(y) be z-independent. Thus functions $f^{(j)}(x,z)$ and $g^{(j)}(y,z)$ are assumed to be stationary solutions of the Schroedinger equation, namely

$$f^{(j)}(x,z) = \exp(\lambda x + i\lambda^2 z)\,\tilde{f}^{(j)}(x,\lambda),\quad g^{(j)}(y,z) = \exp(\mu y + i\mu^2 z)\,\tilde{g}^{(j)}(y,\mu), \quad (6)$$

where $\lambda$ and $\mu$ are complex parameters (see below) and functions $\tilde{f}^{(j)}$ and $\tilde{g}^{(j)}$ depend on the particular potentials v and u, respectively. For instance, the pure wall solution is obtained by setting $a_1 = 2k\sqrt{2\sin\vartheta}\,\exp(i\gamma_0 - kT_0)$, $a_2 = 0$, v = u = 0, $\tilde{f}^{(1)} = \tilde{g}^{(1)} = 1$, $\lambda = a + i\alpha$, $\mu = b + i\beta$ and can be expressed as $E = \exp[i(\beta\cos\vartheta - \alpha\sin\vartheta)L]\,E_{NLS}(T,z)$, where $L = y\cos\vartheta - x\sin\vartheta$ is the (longitudinal) coordinate along the wall direction, $a = k\cos\vartheta$, $b = k\sin\vartheta$ ($\vartheta$ being the angle between the wall and the y-axis) and $E_{NLS} = k\sqrt{2\sin\vartheta}\,\exp\{i[ST/2 - z(\alpha^2 + \beta^2 - k^2) + \gamma_0]\}/\cosh[k(T - T_0 - Sz)]$ is the standard expression of the NLS soliton in the transverse coordinate T, while $S = 2(\alpha\cos\vartheta + \beta\sin\vartheta)$ is the tangent of the angle between the z-axis and the propagation direction, and $\gamma_0$ and $T_0$ are arbitrary real parameters. As is well-known, a characteristic feature of the one-dimensional solitons is that their width and amplitude are related to each other, in contrast to the case of dromions (see below).

As mentioned above, ball-like solutions necessitate of nonvanishing potentials v and u. With the purpose of analytically solving the corresponding Schroedinger equation, we make two different choices of potentials, and show that the main features of the processes we describe are not strongly dependent on the potential. The best known dromion solution is obtained with $a_1 = 0$, $a_2 = A$ (arbitrary complex constant), $v = -2p^2/\cosh^2(px)$, $u = -2q^2/\cosh^2(qy)$ (p and q being positive real parameters) and reads E=Aexp[i($p^2+q^2$)z]/{4coshpx coshqy [1+$|A|^2$(1+tghpx)(1+tghqy)/(64pq)]}. A second simple example of a dromion solution is obtained for binding potentials that are Dirac distributions, namely $v = -2p\,\delta(x)$, $u = -2q\,\delta(y)$; in this case the solution is E= Aexp[i($p^2+q^2$)z]exp($-p|x|-q|y|$)/[1+($|A|^2$/4pq)M(x,p)M(y,q)], with M(x,p)=1−½ exp(−2px) if x>0, M(x,p)= ½ exp(2px) if x<0. The level plot of |E| in the (x,y) plane for both these dromions is shown in Fig.1.

As a direct consequence of the non vanishing external potentials v and u, the wall gets warped and its expression is modified in the neighborhood of the origin x = y =0 with respect to the expression we have given above for v = u =0. This effect can be illustrated by explicitly computing the wall solution for both choices of potentials v and u introduced above. The corresponding level plot is shown in Fig.2 along with some cross-section profiles. These show that, at a large distance from the origin, the wall is an NLS soliton in the transverse coordinate T. In fact, as the longitudinal coordinate L goes to $\pm\infty$, these asymptotic solitons turn out to be merely shifted, with respect to each other, both in phase and position. Thus their expression is again the one given above but with the free parameters $\gamma_0$ and $T_0$ replaced by $\gamma_\pm$ and, respectively, by $T_\pm$ (for L going to $\pm\infty$). The shifts $\Delta T = T_+ - T_-$ and $\Delta\gamma = \gamma_+ - \gamma_-$ turn out to have a simple explicit expression which is easily read out of the following relations: $\exp(-k\Delta T + i\Delta\gamma) = [(\lambda+p)(\mu-q)/(\lambda-p)(\mu+q)]$ for the first choice of the potentials (i.e. $v = -2p^2/\cosh^2(px)$, $u = -2q^2/\cosh^2(qy)$) and $\exp(-k\Delta T + i\Delta\gamma) = [1-(q/\mu)]/[1-(p/\lambda)]$ for potentials $v = -2p\delta(x)$, $u = -2q\delta(y)$. For a completely symmetric set up in the coordinates x and y, say p=q and $\lambda = \mu$, the shifts of phase and position vanish.

We are now in a position to discuss the solution of the DS eq.(2) which describes interaction between ball and wall that we have separately discussed above. This hybrid solution is obtained by inserting functions $f^{(j)}$ and $g^{(j)}$ (see eq.(6)) in the general formula of eq.(3) and by performing the integrals of eq.(5), constructing matrix **A** of eq.(4). Integrals of eq.(5) can be analytically computed only for the Dirac distribution potentials (i.e. for our second choice of potentials above), whereas the asymptotic expressions for very large z can be derived for both choices of external potentials. While we omit detailed computations, we report the main properties of these solutions. We first note that, since the binding potentials v and u do not "move" with z, also the dromion remains confined around the origin x = y =0 for all values of z, for a needle that enters perpendicularly to the (x,y) plane and propagates along the z-direction. On the contrary, the wall moves in the (x,y) plane perpendicularly to its longitudinal direction with rate S= $2(\alpha\cos\vartheta + \beta\sin\vartheta) = 2(\alpha a + \beta b)/k$, which we assume to be non positive, S≤0. In particular, S =0 is of special interest, this being the case in which the stripe and the needle are parallel to each other. In such a situation, the intensity $|E|^2$ is a *periodic* function, depending on z only through the expressions $\sin(\eta z)$ and $\cos(\eta z)$, where $\eta = \alpha^2 + \beta^2 - a^2 - b^2 + p^2 + q^2$. This solution of the DS equation behaves, therefore, like a *breather* and its contour plot in the (x,y) plane is shown in Fig. 3. The oscillation frequency $\eta$ depends on the potentials v and u only through their bound state energy $p^2$ and $q^2$, respectively, and therefore the expression of $\eta$ is the same for both the choices of potential.

Finally, consider the collision between the needle and the stripe, with S<0, shown in Fig.4. For very large and negative values of z, the wall is far away from the origin and its shape is not affected by the dromion (see Fig.4a); indeed it looks like the pure wall solution, with a NLS soliton cross section profile. It then hits the dromion, goes through it, and separates again as z becomes large and positive, while its cross section asymptotically recovers its pure NLS soliton profile. The only effect of the collision on the wall is a shift of the phase, $\Delta\gamma_c$, and of the position, $\Delta T_c$, which is expressed by the relation $\exp(-k\,\Delta T_c + i\,\Delta\gamma_c) = [(\lambda-p)(\mu-q)/(\lambda+p)(\mu+q)]$ for the potentials $v = -2p^2/\cosh^2(px)$, $u = -2q^2/\cosh^2(qy)$, and $\exp(-k\,\Delta T_c + i\,\Delta\gamma_c) = [1-(p/\lambda)][1-(q/\mu)]$ for the other choice of the potentials, $v = -2p\,\delta(x)$, $u = -2q\,\delta(y)$. As for the ball, its shape changes considerably as a consequence of the collision, as is quite apparent in Fig.4. Moreover, the total energy, $H = \int_{-\infty}^{+\infty}dx \int_{-\infty}^{+\infty}dy\,|E|^2$, although conserved, is *infinite* because of the wall component, and a conclusion about the conservation of the dromion energy cannot be directly derived from the break-up process we have described. We conjecture that its energy does not change, given that a shift of the phase and position of the cross section profile of the wall does not change its energy density. In addition, this conclusion agrees with the similar dromion energy conservation found in the dromion-dromion collision [9], if the constant matrix **R** (see the definition of eq.(4)) is diagonal, as it is in the present case.

In conclusion we have studied analytically the hybrid collision of a one-dimensional stripe soliton and a two-dimensional needle in the framework of the Davey-Stewartson equation.

The work by E.D. was carried out in the framework of an agreement between Fondazione Ugo Bordoni and the Italian Communications Administration. A.D. gratefully acknowledges support from the Italian Ministero della Universita' e della Ricerca Scientifica e Tecnologica and the Mexican Centro Internacional de Ciencias (CIC) in Cuernavaca, and thanks Thomas Seligman and Carlos Mejia Monasterio for their assistance.

**Figure Captions**

**Fig.1:** Level plot of $|E|$ for the dromion solution, with p=1, q=2, and A=4, for (a) $v = -2p^2/\cosh^2(px)$, $u = -2q^2/\cosh^2(qy)$ and (b) $v = -2p\,\delta(x)$, $u = -2q\,\delta(y)$.

**Fig.2:** (a) Level plot of $|E|$ for the wall solution for the values $p=1$, $q=2$, $a_1=4$, $\lambda=1.01+2i$ and $\mu=2.1+0.5i$; (b-e) corresponding sections for L=4, 2, -2 and -4 respectively, as a function of T.

**Fig.3:** Level plot of $|E|$ for the dromion-wall solution (with Dirac delta function potentials) for the values $p=1$, $q=2$, $\lambda=2$, $\mu=3$, $a_1=4$ and $a_2=0.04$ for which $S=0$. The six pictures are obtained respectively for (2πn/5η) with n=0,1,2,3,4,5. The period is $2\pi/|\eta|=\pi/4$.

**Fig.4:** Level plot of $|E|$ for the dromion-wall solution (with Dirac delta function potentials) for parameter values $p=1$, $q=2$, $\lambda=1.01-2i$, $\mu=2.1+0.5i$, $a_1=4$ and $a_2=6$, (a) at z=-8 and (b) at z=8.

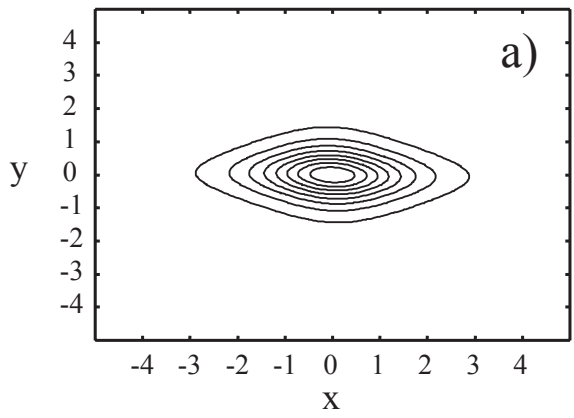 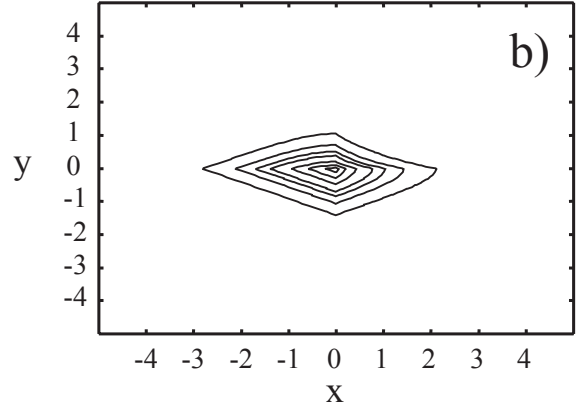

Figure 1 Ciattoni et al.

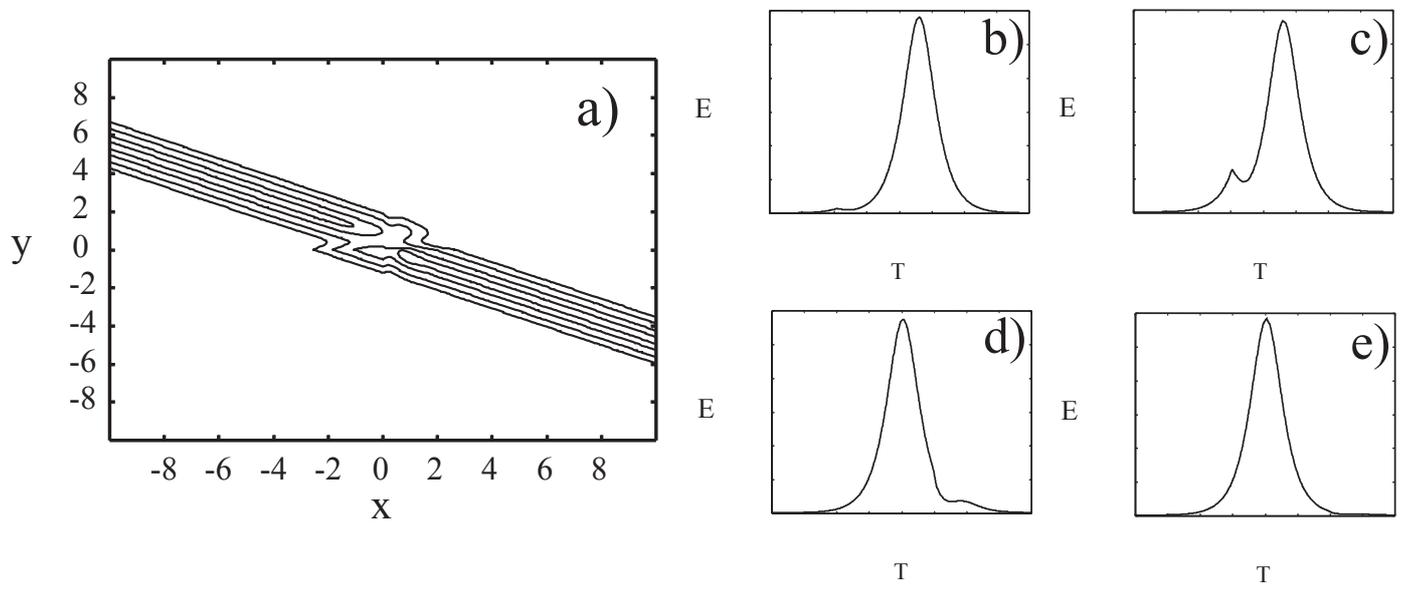

Figure 2 Ciattoni et al.

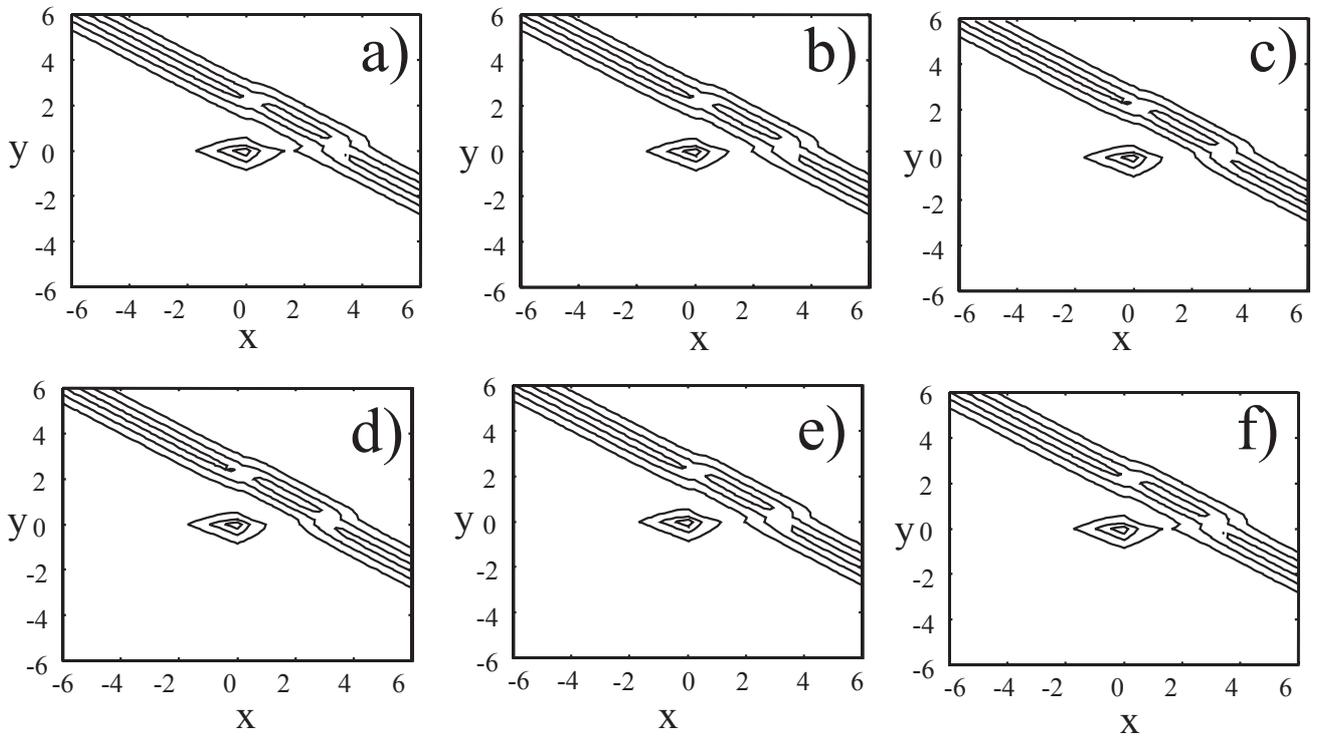

Figure 3 Ciattoni et al.

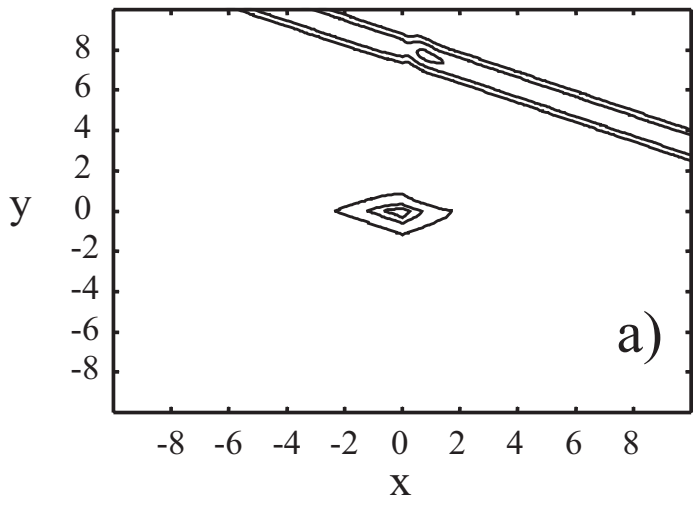 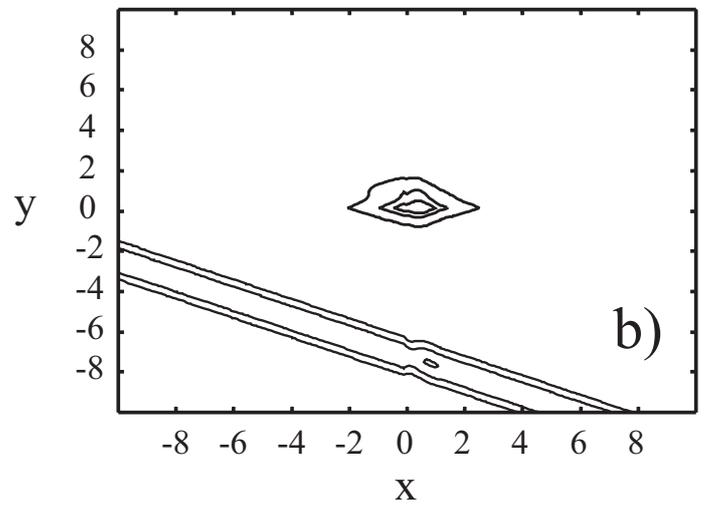

Figure 4 Ciattoni et al.